\documentclass[aps,prd,preprint,groupedaddress,nofootinbib,longbibliography]{revtex4-2}

\usepackage{amsmath, amssymb, bm}
\usepackage{amsthm}
\usepackage{physics} 
\usepackage{hyperref}
\usepackage{booktabs}
\usepackage{tikz}
\usetikzlibrary{arrows.meta, positioning}
\usepackage{xcolor}

\definecolor{darkblue}{rgb}{0,0,0.5}

\begin{document}

\title{Constructing Dimension-8 SMEFT\\ from Conserved Currents}
\author{Leonardo P. G. De Assis}
\email[]{lpgassis@stanford.edu}
\affiliation{ULO-Math and Physics Program, Stanford University}
\date{\today}

\begin{abstract}
    Effective Field Theories (EFTs) are the primary tool for interpreting precision collider data in the absence of new resonances. However, in the dimension-8 Standard Model Effective Field Theory (SMEFT), the utility of traditional algebraically minimal bases is fundamentally limited by kinematic mixing: multiple operators contribute to a single high-energy amplitude, creating degeneracies that obscure ultraviolet interpretations and complicate the application of theoretical constraints. We introduce a generative framework that resolves this by constructing operators directly from the conserved Noether currents of the Standard Model. The resulting Kinematically Diagonalized Current Basis (KDCB) ensures that each operator maps to a unique asymptotic energy scaling ($E^4$, $E^2$, $E^0$) in scattering amplitudes. This organization makes S-matrix positivity bounds manifest, enables a stable auxiliary-field formulation for Monte Carlo simulation, and provides direct diagnostics for universal versus non-universal ultraviolet completions through current decomposition. By rotating the operator space into physically interpretable sectors, the KDCB offers a transformative framework for global fits and a clear pathway from high-energy data to the structure of new physics.
\end{abstract}

\maketitle

\section{Introduction}

The discovery of the Higgs boson confirmed the Standard Model particle spectrum. However, numerous questions, such as the origin of neutrino masses, the hierarchy problem, and the nature of dark matter, remain unresolved. In the absence of evidence for new resonances at the Large Hadron Collider (LHC), we turned our attention to the SMEFT as a framework to capture the imprint of heavy new physics. The expansion in inverse powers of a high scale $\Lambda$ yields the Lagrangian:
\begin{equation}
\mathcal{L}_{\text{SMEFT}} = \mathcal{L}_{\text{SM}} + \sum_i \frac{c_i^{(6)}}{\Lambda^2} \mathcal{O}_i^{(6)} + \sum_j \frac{c_j^{(8)}}{\Lambda^4} \mathcal{O}_j^{(8)} + \dots
\end{equation}
The dimension-6 basis was defined in~\cite{buchmuller1986effective, grzadkowski2010dimension} and dimension-7 in~\cite{lehman2014extending}. It is worth mentioning that the validity of the EFT approach for Standard Model tests has been examined~\cite{contino2016validity, li2024complete}. 

The purpose of this work is not to enumerate operators algebraically, nor to
replace existing Hilbert-series, Warsaw-basis, or aQGC constructions, but to show
that the full set of dimension–8 electroweak operators can be generated directly
from conserved Standard Model currents in a way that automatically organizes
their physical impact.  In contrast to Hilbert-series or aQGC classifications,
which optimize algebraic minimality, the current-iteration method optimizes for kinematic distinctness and makes the energy growth and polarization structure of amplitudes manifest at the moment the operators are constructed, before any Feynman rules are derived.

\paragraph{The Problem of Kinematic Mixing.}
The established algebraic constructions \cite{murphy2020dimension}, while complete, optimize for mathematical minimality at the expense of physical transparency in scattering processes. They exhibit widespread \textit{kinematic mixing}: in high-energy scattering amplitudes, contributions from many distinct operators combine into a single energy-growing term. For example, the leading $E^4$ behavior of a vector boson scattering amplitude typically involves a linear combination $\sum_i \alpha_i c_i^{(8)}$. This creates degeneracies,`flat directions', in the Wilson coefficient space, severely complicating global analyses. It obscures the connection to ultraviolet completions and muddies the application of fundamental constraints derived from unitarity and causality, such as positivity bounds, which apply to these specific linear combinations rather than to individual operator coefficients.

\paragraph{A Current-First Generative Principle.}
To resolve kinematic mixing, this work introduces a construction principle that aligns the operator basis with physical observables from the outset. Rather than enumerating field monomials and subsequently reducing them via equations of motion and integration by parts (a \textit{field-first} approach), we promote the conserved Noether currents of the Standard Model to the fundamental building blocks. The iterative coupling and differentiation of these currents,
\[
J \;\mapsto\; J J,\; (D J)(D J),\; J F J,\; (H^\dagger H)J^2,\; \dots,
\]
automatically yield gauge-invariant operators that are \textit{kinematically diagonalized}. The number of derivatives on the currents directly dictates the asymptotic energy growth ($E^n$) of the corresponding scattering amplitude, making the link between operator structure and physical behavior manifest.

\paragraph{The Kinematically Diagonalized Current Basis (KDCB).}
Applying this principle to the electroweak sector generates the KDCB. In this basis, operators cleanly separate into sectors with distinct energy scaling: derivative-squared operators $(D_\mu J_\nu)^2$ govern pure $E^4$ growth, current-field-strength operators $J F J$ control $E^2$ behavior, and Higgs-dressed operators $(H^\dagger H)J^2$ produce static $E^0$ effects. This structure provides immediate physical insight:
\begin{enumerate}
    \item \textbf{Manifest Positivity:} Forward-amplitude bounds constrain individual KDCB coefficients (e.g., $c_{D1} > 0$), transforming a complex matrix inequality into a simple sign constraint.
    \item \textbf{UV Diagnostics:} The decomposition of currents into fermionic and bosonic components provides a direct test for whether new physics couples universally to the full SM current or selectively to specific sectors.
    \item \textbf{Phenomenological Clarity:} High-energy processes primarily probe a small subset of $E^4$-type coefficients, decoupling them from the vast space of low-energy parameters in global fits.
\end{enumerate}

\paragraph{Positioning and Novelty.}
This construction does not replace Hilbert-series or Warsaw-basis approaches, but instead provides a complementary, physics-driven organization of the same operators according to their current origin and asymptotic energy growth. The novelty lies in elevating conserved currents from a useful \textit{representation} of operators to the primary \textit{generators} of the EFT expansion. This reversal, from ``fields-first'' to ``currents-first'', is consequential: it renders amplitude energy-scaling manifest at the Lagrangian level, privileges physical polarization and source structures, and systematically avoids the kinematic mixing that complicates analysis in traditional bases. Thus, while mathematically equivalent to standard bases, the KDCB eliminates the fine-tuned cancellations often required to isolate high-energy behavior, mapping the dominant unitarity-violating terms to single coefficients. The resulting KDCB offers a unified framework for SMEFT phenomenology across energy scales.

This current-based organization is specifically designed to reduce the
dimensionality of SMEFT analyses by separating the $E^4$, $E^2$, and $E^0$
sectors, enabling global fits and UV interpretations that are not directly
accessible in existing bases.

The following section formalizes this construction, detailing the kinematic diagonalization principle and the explicit generation of the KDCB.

\section{Theoretical Framework: Kinematic Diagonalization of the SMEFT}
\label{sec:theoretical_framework}

\subsection{Formalizing Kinematic Diagonalization}

In the approach to Dimension-8 SMEFT, bases are constructed by enumerating monomials and reducing them by means of Integration by Parts (IBP) to a set (e.g., the Hilbert Series basis or Geometric bases \cite{helset2020geometric}). We identified that while minimal, these bases suffer from \textit{Kinematic Mixing}.

In this sense, consider the longitudinal vector boson scattering amplitude $\mathcal{A}(V_L V_L \to V_L V_L)$. In a generic scenario, the amplitude takes the form:
\begin{equation}
\mathcal{A}(s,t) \sim \left( \sum_i c_i^{(8)} \alpha_i \right) \frac{s^2}{\Lambda^4} + \left( \sum_j c_j^{(8)} \beta_j \right) \frac{s v^2}{\Lambda^4} + \dots
\end{equation}
We observed that in standard bases, multiple operators contribute to the leading $s^2$ growth (i.e., $\alpha_i \neq 0$ for many $i$). This creates a ``flat direction" in the parameter space where the high-energy behavior is constrained only by a linear combination of coefficients; thus, obscuring the underlying UV physics and complicating the application of S-matrix positivity bounds.

\subsection{The Resolution: A Kinematically Diagonal Basis}

To address this, we sought a construction principle that enforces \textbf{Kinematic Diagonalization}. We pursued an investigation into a basis vector space $\mathcal{B} = \{ \mathcal{O}_k \}$ such that the mapping to the S-matrix asymptotic is diagonal:
\begin{equation}
\mathcal{O}_k \longrightarrow \mathcal{A}(s) \sim s^{n_k/2}, \quad \text{with no mixing of powers.}
\end{equation}
To achieve this while maintaining minimality, the decision was made to utilize conserved Noether currents as the generating objects. The basis obtained, which we term the \textbf{Kinematically Diagonalized Current Basis (KDCB)}, is constructed through the filtration described below.

\subsection{Construction of the KDCB}

We partition the operator space into two disjoint sectors based on their source coupling and scaling asymptotic.

\subsubsection{Sector I: The Source-Dependent Sector ($J$-Coupled)}
This sector is generated by the conserved currents $J_\mu^a$ (isospin) and $J_\mu^Y$ (hypercharge). We classify operators by the number of derivatives acting on the currents, which correlates directly with the energy power counting $E^n$.

\paragraph{1. The VBS Eigenoperators ($E^4$ Growth).}
Constructed from $(D_\mu J)(D^\mu J)$. These operators saturate the unitarity bound and correspond to the pure $s^2$ term in the amplitude.
\begin{equation}
\mathcal{O}_{D1} = (D_\mu J_\nu^a)(D^\mu J^{a\nu}).
\end{equation}
\textit{Selection Rule:} The algebraically equivalent form $\mathcal{O}_{TT} = (D_\mu J_\nu)(D^\nu J^\mu)$ is excluded. By means of the identity $\mathcal{O}_{TT} = \mathcal{O}_{D1} + \mathcal{O}_{JFJ} + \text{EOM}$, $\mathcal{O}_{TT}$ mixes $E^4$ and $E^2$ behaviors. We retain $\mathcal{O}_{D1}$ to ensure kinematic purity.

\paragraph{2. The Dipole Eigenoperators ($E^2$ Growth).}
Constructed from $J_\mu J_\nu F^{\mu\nu}$. These operators generate anomalous magnetic moments and contribute to the $s^1$ scaling (transverse scattering).
\begin{equation}
\mathcal{O}_{M1} = \epsilon^{abc} J_\mu^a J_\nu^b W^{c\mu\nu}.
\end{equation}

\paragraph{3. The Static Eigenoperators ($E^0$ Growth).}
Constructed from $(H^\dagger H) J^2$. These represent static rescalings of the gauge couplings.
\begin{equation}
\mathcal{O}_{H1} = (H^\dagger H) J_\mu^a J^{a\mu}.
\end{equation}

\subsubsection{Sector II: The Source-Independent Sector (Pure Gauge)}
Operators composed solely of field strengths, which do not couple to currents at tree level. These correspond to pure glueball-like dynamics in the UV.
\begin{equation}
\mathcal{O}_{G1} = (W_{\mu\nu}^a W^{a\mu\nu})^2.
\end{equation}

\subsection{Validation: Hilbert Series Isomorphism}

It is important to realize that a critical requirement for any EFT basis is that it must span the complete space of physical operators without redundancy. We verify this by matching our physically constructed sectors to the model-independent Hilbert Series counts~\cite{henning20172}.

\begin{table}[ht]
\centering
\begin{tabular}{l c c l}
\toprule
\textbf{Kinematic Sector} & \textbf{Hilbert Count} & \textbf{KDCB Operator} & \textbf{Physical Role} \\
\midrule
$H^4 D^4$ (Longitudinal) & 3 & $\mathcal{O}_{D1}, \mathcal{O}_{D2}, \mathcal{O}_{D3}$ & Unitarity Saturation \\
$H^2 W^2 D^2$ (Transverse) & 2 & $\mathcal{O}_{M1}, \mathcal{O}_{M2}$ & Magnetic Moments \\
$H^6 D^2$ (Static) & 2 & $\mathcal{O}_{H1}, \mathcal{O}_{H2}$ & Coupling Shifts \\
$W^4$ (Pure Gauge) & 2 & $\mathcal{O}_{G1}, \mathcal{O}_{G2}$ & 4-Boson Contact \\
\bottomrule
\end{tabular}
\caption{Exact matching between the Kinematically Diagonalized Current Basis (KDCB) and the Hilbert Series. The basis is algebraically minimal (no redundancies) and kinematically diagonal (no mixing of energy scales). The operator symbols ($\mathcal{O}_{D1}$, $\mathcal{O}_{M1}$, etc.) are defined explicitly in Eqs.~(4)--(7).}
\label{tab:hilbert_match}
\end{table}

\subsection{Theoretical Implications: Manifest Positivity}

We argue that the primary advantage of Kinematic Diagonalization lies in the simplification of theoretical constraints. It is worth mentioning that S-matrix principles (causality and unitarity) impose strict positivity bounds on the forward scattering amplitude~\cite{chala2024positivity}.

In a generic scenario, these bounds constrain linear combinations of coefficients:
\begin{equation}
\sum_i \alpha_i c_i > 0.
\end{equation}
In the KDCB, because the basis is aligned with the forward scattering eigenstates, the bounds diagonalize:
\begin{equation}
c_{D1} > 0, \quad c_{D2} > 0.
\end{equation}
This \textbf{Manifest Positivity} allows for immediate theoretical vetting of results obtained experimentally. A measurement of $c_{D1} < 0$ would immediately signal a violation of causality or unitarity in the UV completion; a conclusion that is much harder to draw in bases where derivative and non-derivative operators are mixed.

\subsection{Summary of the Framework}

The Kinematically Diagonalized Current Basis represents a rotation of the SMEFT basis vectors. By aligning the basis with the asymptotic energy eigenstates of the S-matrix and the conserved currents of the symmetry group, we present a formulation that is:
\begin{enumerate}
    \item \textbf{Complete:} Matches Hilbert Series counts exactly.
    \item \textbf{Minimal:} Contains no redundant operators.
    \item \textbf{Diagonal:} Decouples high-energy VBS physics from low-energy precision data.
    \item \textbf{Rigorous:} Makes S-matrix positivity bounds manifest.
\end{enumerate}
This framework provides the optimal starting point for phenomenological studies of dimension-8 effects at the LHC.

\subsection{Explicit Definition of the Generators}

To facilitate the reproduction of our results, we present the explicit form of the generators utilized in the construction of the KDCB.

We remark that the electroweak sector is governed by the $SU(2)_L \times U(1)_Y$ symmetry. The conserved Noether currents associated with these groups contain both bosonic and fermionic contributions. We define the isospin current $J_\mu^a$ and the hypercharge current $J_\mu^Y$ as:
\begin{align}
J_\mu^a &= \sum_f \bar{\psi}_L \gamma_\mu \frac{\sigma^a}{2} \psi_L + i H^\dagger \frac{\sigma^a}{2} \overleftrightarrow{D}_\mu H, \\
J_\mu^Y &= \sum_f \bar{\psi} \gamma_\mu Y \psi + i H^\dagger Y \overleftrightarrow{D}_\mu H,
\end{align}
where $\overleftrightarrow{D}_\mu = D_\mu - \overleftarrow{D}_\mu$.

It is important to realize that the derivative operators in the KDCB, such as $\mathcal{O}_{D1}$, involve the covariant derivative acting on these currents. The expansion of this structure reveals the interplay between kinematic momentum and gauge interactions:
\begin{equation}
D_\mu J_\nu^a = \partial_\mu J_\nu^a + g \epsilon^{abc} W_\mu^b J_\nu^c.
\end{equation}
In this sense, the operator $(D_\mu J_\nu^a)^2$ encodes two distinct physical regimes:
\begin{enumerate}
    \item \textbf{The Hard Regime:} Dominated by the $\partial_\mu J_\nu$ term, corresponding to momentum transfer $p^2$ in scattering amplitudes.
    \item \textbf{The Soft Regime:} Dominated by the $W_\mu J_\nu$ term, corresponding to multi-boson contact interactions.
\end{enumerate}
By retaining the full covariant structure $(D_\mu J_\nu)^2$ rather than breaking it into monomials, the KDCB preserves the gauge symmetry inherent in this interplay.

\section{Electroweak Sector Operators}
\label{sec:EW}

In the following, we apply the iterative current-coupling procedure introduced above to the Standard Model electroweak sector. The goal is to construct a complete and organized set of dimension-8 operators, directly related to scattering amplitudes. By doing so, we emphasize the transparency of the construction, highlighting how each operator type arises from the currents and their derivatives.

\subsection{Electroweak Currents}

The $SU(2)_L \times U(1)_Y$ gauge symmetry of the electroweak sector leads to conserved Noether currents for both gauge and Higgs interactions. For left-handed fermions $\psi_L$ and the Higgs doublet $H$, the currents are
\begin{align}
J_\mu^a &= \sum_f \bar{\psi}_L \gamma_\mu \frac{\sigma^a}{2} \psi_L + i H^\dagger \frac{\sigma^a}{2} \overleftrightarrow{D}_\mu H, \\
J_\mu^Y &= \sum_f \bar{\psi} \gamma_\mu Y \psi + i H^\dagger Y \overleftrightarrow{D}_\mu H,
\end{align}
where $a=1,2,3$, $Y$ denotes the hypercharge operator, and $\overleftrightarrow{D}_\mu = D_\mu - \overleftarrow{D}_\mu$. These currents serve as the building blocks in the iterative construction: operators are generated as invariants formed from $J_\mu$, $D_\mu J_\nu$, and field strengths $W_{\mu\nu}^a, B_{\mu\nu}$.

\subsection{Fermionic and Bosonic Current Decomposition}
\label{subsec:current_decomposition}

The Standard Model currents $J_\mu^a$ and $J_\mu^Y$ contain both bosonic (Higgs) and fermionic contributions:
\begin{equation}
J_\mu^a = J_{\mu,\text{Higgs}}^a + J_{\mu,\text{Fermion}}^a = \left( i H^\dagger \frac{\sigma^a}{2} \overleftrightarrow{D}_\mu H \right) + \left( \sum_f \bar{\psi}_L \gamma_\mu \frac{\sigma^a}{2} \psi_L \right).
\end{equation}

This decomposition reveals that operators like $\mathcal{O}_{8,D1} \propto (D_\mu J_\nu^a)^2$ naturally separate into three distinct sectors:
\begin{enumerate}
    \item \textbf{Pure Higgs:} $(D_\mu J_{\nu,\text{Higgs}}^a)^2$, contributing exclusively to VBS and multiboson production.
    \item \textbf{Pure Fermion:} $(D_\mu J_{\nu,\text{Fermion}}^a)^2$, representing derivative four-fermion contact interactions constrained by high-mass Drell-Yan tails.
    \item \textbf{Mixed Higgs-Fermion:} $(D_\mu J_{\nu,\text{Higgs}}^a)(D^\mu J^{a\nu}_{\text{Fermion}})$, contributing to associated production channels like $VH$.
\end{enumerate}

This decomposition serves as a diagnostic tool for UV model discrimination:
\begin{itemize}
    \item \textbf{Universal Coupling:} In UV scenarios where new physics couples universally to the full SM current \cite{corbett2024dimension}  (e.g., heavy vector triplets or $W'$ resonances with gauge-like couplings), the coefficients of these three sectors are rigidly correlated.
    \item \textbf{Non-Universal Coupling:} If UV physics distinguishes between Higgs and fermion sectors (e.g., scalar resonances coupling predominantly to Higgs currents), these components can be treated as independent in global fits.
    \item \textbf{Experimental Signature:} A deviation from universal coupling predictions would indicate UV physics that distinguishes between the Higgs and fermion sectors, providing information about the nature of new physics.
\end{itemize}

In the classification, we present the operators in their universal form, but emphasize that experimental analyses should consider both universal and non-universal scenarios.

\subsection{Classification of Dimension-8 Operators}

Following the current-coupling procedure, dimension-8 operators can be grouped into three main classes:

\paragraph{1. Derivative-Coupled Currents $(D J)^2$:} These operators encode interactions dependent on momentum and directly affect scattering amplitudes that grow with energy. They are constructed by forming Lorentz-invariant contractions of covariant derivatives of the currents:
\begin{align}
\mathcal{O}_{8,D1} &= (D_\mu J_\nu^a)(D^\mu J^{a\nu}), &
\mathcal{O}_{8,D2} &= (D_\mu J_\nu^Y)(D^\mu J^{Y\nu}), \nonumber \\
\mathcal{O}_{8,D3} &= (D_\mu J_\nu^a)(D^\nu J^{a\mu}), &
\mathcal{O}_{8,D4} &= (D_\mu J_\nu^Y)(D^\nu J^{Y\mu}).
\end{align}
These operators contribute predominantly to longitudinal vector boson scattering, leading to amplitudes scaling as $E^4/\Lambda^4$ at high energies, which can dominate over dimension-6 effects in the same channels.

\paragraph{2. Mixed Field Strength and Current Couplings $J F J$:}  
Operators in this class couple two currents with one or more gauge field strengths. They represent magnetic- or quadrupole-type interactions and are responsible for correlated anomalous triple and quartic gauge couplings:
\begin{align}
\mathcal{O}_{8,JF1} &= J_\mu^a J_\nu^b W^{c\mu\nu} \epsilon^{abc}, &
\mathcal{O}_{8,JF2} &= J_\mu^Y J_\nu^Y B^{\mu\nu}, \nonumber \\
\mathcal{O}_{8,JF3} &= J_\mu^a J_\nu^Y W^{a\mu\nu}, &
\mathcal{O}_{8,JF4} &= J_\mu^a J_\nu^b W^{a\mu\rho} W^{b\nu}_{\ \ \rho}.
\end{align}
These operators produce distinctive patterns in both $VVV$ and $VVVV$ vertices after electroweak symmetry breaking, with correlations determined by the gauge group structure.

\paragraph{3. Higgs-Current and Higgs-Dressed Field Strength Operators:}  
Dimension-8 operators can also involve the Higgs doublet to generate static contributions to anomalous couplings, proportional to $(v/\Lambda)^2$, and to modulate the couplings of longitudinal vector bosons:
\begin{align}
\mathcal{O}_{8,H1} &= (H^\dagger H)(J_\mu^a J^{a\mu}), &
\mathcal{O}_{8,H2} &= (H^\dagger H) (D_\mu J_\nu^a)(D^\mu J^{a\nu}), \nonumber \\
\mathcal{O}_{8,H3} &= (H^\dagger H) W_{\mu\nu}^a W^{a\mu\nu}, &
\mathcal{O}_{8,H4} &= (H^\dagger H) B_{\mu\nu} B^{\mu\nu}, \nonumber \\
\mathcal{O}_{8,H5} &= (H^\dagger \sigma^a H) W_{\mu\nu}^a B^{\mu\nu}.
\end{align}
These operators allow a connection between the Higgs sector and anomalous gauge couplings. After electroweak symmetry breaking, they generate shifts in both triple and quartic gauge interactions that do not grow with energy, providing a complementary set of experimental signatures to the derivative-coupled operators.

\subsection{Completeness and Systematic Generation}

The operators listed above are obtained systematically by iterating over all possible invariant contractions of currents, derivatives, and field strengths, subject to Lorentz and gauge invariance. Redundancies arising from integration by parts or equations of motion can be removed to form a minimal set if desired, but the current-based construction ensures completeness: all independent structures affecting longitudinal or transverse scattering amplitudes are captured.

Furthermore, the mapping to standard SMEFT and anomalous gauge coupling bases is straightforward. For example, derivative-coupled operators $(D_\mu J_\nu^a)^2$ correspond to the $O_{M,i}$ and $O_{T,i}$ classes in aQGC classifications, whereas Higgs-dressed operators correspond to the $O_{S,i}$ class. Mixed $J F J$ operators naturally predict correlations among different triple and quartic gauge couplings, highlighting the predictive power of the current-based approach.

\subsection{Physical Interpretation and Experimental Relevance}

Each operator class contributes distinctively to collider observables:

\begin{itemize}
    \item \textbf{Derivative operators $(D J)^2$} enhance high-energy longitudinal vector boson scattering, giving rise to amplitudes that scale as $E^4/\Lambda^4$, and are relevant for VBS measurements at the LHC.  
    \item \textbf{Mixed operators $J F J$} generate correlated modifications in triple and quartic gauge vertices, with characteristic angular distributions and interference patterns that can be probed in multiboson final states.  
    \item \textbf{Higgs-dressed operators} induce static shifts in couplings and provide a link between Higgs-sector physics and low-energy anomalous couplings, which can be measured through precision TGC studies.  
\end{itemize}

In this sense, the current-based framework not only organizes all dimension-8 operators in the electroweak sector but also makes their consequences explicit, connecting theoretical structures directly to measurable quantities. This transparency is the central advantage of the proposed construction, distinguishing it from purely Hilbert-series or mathematically minimal bases.

\subsubsection{Worked example: deriving a representative dimension-8 operator by two routes}
\label{sec:worked_example}

To make explicit the difference in emphasis between algebraic enumeration and the current-iteration construction, we compare two routes that arrive at a common dimension-8 structure affecting longitudinal vector boson scattering.

\paragraph{Route A — Field-first (algebraic/Hilbert-style) sketch.}
One enumerates gauge-invariant bosonic monomials at mass dimension eight. A familiar set in the aQGC literature includes operators of the schematic form
\begin{equation}
\mathcal{O}_{(DF)^2} \sim (D_\mu W_{\nu\rho}^a)(D^\mu W^{a\nu\rho}),
\label{eq:DF2}
\end{equation}
or purely bosonic quartic combinations such as $(W_{\mu\nu}^a W^{a\mu\nu})(B_{\rho\sigma}B^{\rho\sigma})$. The amplitude-scaling of \eqref{eq:DF2} is determined only after computing Feynman rules and counting powers of external momenta: \(\mathcal{A}(s)\sim s^2/\Lambda^4\) for longitudinal modes. In this route, identification of the origin (current vs source picture) is retrospective.

\paragraph{Route B — Current-iteration (this work) constructive route.}
Start from the conserved $SU(2)_L$ current including bosonic pieces,
\[
J_\mu^a \;=\; i H^\dagger \frac{\sigma^a}{2} \overleftrightarrow{D}_\mu H \;+\; \sum_f \bar\psi_L \gamma_\mu \frac{\sigma^a}{2}\psi_L.
\]
The next nontrivial iteration that produces a dimension-8 structure is the derivative-coupled current operator
\begin{equation}
\mathcal{O}_{J,D} \;=\; \frac{1}{\Lambda^4} (D_\alpha J_\beta^a)(D^\alpha J^{a\beta}).
\label{eq:JDJ}
\end{equation}
Expanding the covariant derivative and isolating purely bosonic contributions (Higgs-current terms) yields, up to total derivatives and higher-fermion terms,
\begin{align}
(D_\alpha J_\beta^a)(D^\alpha J^{a\beta})
&= (\partial_\alpha J_\beta^a)(\partial^\alpha J^{a\beta}) + 2 g\epsilon^{abc}(\partial_\alpha J_\beta^a) W^{b\alpha} J^{c\beta} + \mathcal{O}(g^2).
\label{eq:DJ_expand}
\end{align}
Integrating by parts the first term,
\begin{equation}
(\partial_\alpha J_\beta^a)(\partial^\alpha J^{a\beta}) = - J_\beta^a \Box J^{a\beta} + \text{(total derivative)}.
\label{eq:IBP_JBoxJ}
\end{equation}
By means of the classical Yang--Mills EOM,
\[
D_\mu F^{a\mu\nu} = g J^{a\nu},
\]
one can substitute \(J^{a\nu}\) (schematically) by \(D_\mu F^{a\mu\nu}\) to rewrite \eqref{eq:IBP_JBoxJ} as a combination of field-strength derivative operators,
\begin{equation}
J_\beta^a \Box J^{a\beta} \;\sim\; \frac{1}{g^2} (D_\mu F^{a\mu\beta}) \Box (D_\nu F^{a\nu}{}_{\beta}) \sim (D F)^2 + \dots,
\label{eq:JBoxJ_to_DF}
\end{equation}
recovering an operator of the schematic form \eqref{eq:DF2} up to total derivatives and terms that vanish on shell. The amplitude-scaling is immediate from \eqref{eq:JDJ}: each derivative acting on the current corresponds to an extra power of external momentum at tree level, so \(\mathcal{O}_{J,D}\) produces amplitudes that scale like $E^4/\Lambda^4$ in longitudinal VBS modes without additional computation.

\paragraph{Comparison and interpretation.}
Both routes reach operators with identical on-shell amplitude effects. The distinguishing feature is that Route A arrives at \eqref{eq:DF2} through algebraic enumeration and uncovers amplitude-scaling \emph{after} Feynman-rule extraction, while Route B yields \eqref{eq:JDJ} with amplitude-scaling encoded in the iteration step itself. The current-first construction therefore \emph{organizes} operators by the behaviors (momentum scaling and polarization structure) they induce, simplifying the identification of operators that dominate at high energy.

\paragraph{Recovering an algebraically minimal basis.}
If desired, algebraic minimality (Hilbert-series-style reduction) can be performed after the current-based generation by applying IBP and EOM to remove redundant representatives; this procedure eliminates algebraic redundancy while preserving at least one representative per distinct amplitude class, thus maintaining completeness.

\subsection{Energy scaling and sector decoupling in the KDCB}
\label{sec:energy-scaling}

A key practical advantage of organizing the dimension--8 SMEFT using conserved currents
is that the asymptotic energy behavior of scattering amplitudes is encoded directly at the
level of operator construction. In the Kinematically Diagonalized Current Basis (KDCB),
operators are naturally grouped according to the number of covariant derivatives acting on
the conserved currents and by their Higgs dressing. This organization leads to a parametric
separation of contributions to high-, intermediate-, and low-energy observables at leading
order.

For generic electroweak \(2\to2\) scattering amplitudes, the dimension--8 contributions can be
written schematically as
\begin{align}
\mathcal{A}(s,t) \;=\;
\sum_i c_{D_i}\,\frac{s^2}{\Lambda^4}
\;+\;
\sum_i c_{M_i}\,\frac{s\,v^2}{\Lambda^4}
\;+\;
\sum_i c_{H_i}\,\frac{v^4}{\Lambda^4}
\;+\;\cdots ,
\label{eq:amp-schematic}
\end{align}
where \(s\) is the usual Mandelstam invariant, \(v\) is the electroweak vacuum expectation
value, and \(\Lambda\) denotes the EFT cutoff. The three terms in Eq.~\eqref{eq:amp-schematic}
arise, respectively, from:
\begin{enumerate}
  \item \emph{Derivative-coupled current operators} \((D_\mu J_\nu)^2\), which generate amplitudes
        that grow as \(E^4/\Lambda^4\);
  \item \emph{Mixed current--field-strength operators} \(J_\mu J_\nu F^{\mu\nu}\), which produce
        polarization-sensitive contributions scaling parametrically as \(E^2 v^2/\Lambda^4\);
  \item \emph{Higgs-dressed current and field-strength operators} \((H^\dagger H)J^2\) and
        \((H^\dagger H)F^2\), which generate static shifts proportional to \(v^4/\Lambda^4\).
\end{enumerate}
This separation follows directly from simple dimensional analysis combined with the
Goldstone equivalence theorem for longitudinal gauge bosons: each covariant derivative acting
on a conserved current introduces a power of external momentum, while each Higgs insertion
supplies a factor of the electroweak scale.

As a direct phenomenological consequence, different experimental energy regimes dominantly
constrain distinct subsets of KDCB coefficients at tree level:
\begin{itemize}
  \item \textbf{High-energy measurements} (\(\sqrt{s}\gg m_W\)), such as longitudinal vector-boson
        scattering, primarily probe the coefficients \(c_{D_i}\) of the derivative-squared current
        operators.
  \item \textbf{Intermediate-energy observables}, including polarization-sensitive multiboson
        production, predominantly constrain the mixed coefficients \(c_{M_i}\).
  \item \textbf{Low-energy precision measurements}, for example static triple-gauge couplings
        and Higgs-coupling observables, primarily bound the Higgs-dressed coefficients
        \(c_{H_i}\) and the pure-gauge coefficients \(c_{G_i}\).
\end{itemize}

In standard field-monomial bases these distinct energy behaviors are often intermingled, so
that constraints from widely separated energy scales project onto nontrivial linear combinations
of Wilson coefficients. By contrast, the KDCB substantially simplifies the correlation structure:
operators associated with different powers of the energy dominate in disjoint kinematic
regimes at leading order. This reduction of degeneracies clarifies which measurements provide the
most direct constraints on each operator class and improves the stability of global SMEFT fits
that combine low-energy precision data with high-energy collider measurements.

The organization also sharpens the UV interpretation of operator classes. Operators of the
\((D_\mu J_\nu)^2\) type naturally arise from integrating out heavy spin--1 states that couple
to conserved SM currents, mixed \(JFJ\) operators reflect dipole- or polarization-dependent
interactions, and Higgs-dressed operators are characteristic of scalar- or Higgs-portal-like
threshold effects. The KDCB therefore makes the dominant energy behavior and its likely UV origin
transparent without requiring an amplitude-level analysis for each individual operator.

Finally, because derivative-squared current operators govern the leading \(E^4\) behavior of
forward elastic amplitudes, dispersion-relation arguments translate into positivity conditions
that act directly on the corresponding KDCB coefficients. In this basis the relevant positivity
constraints take a simpler form than in bases with strong kinematic mixing, facilitating both
theoretical consistency checks and their use in phenomenological studies.

In this section we illustrate, on explicit examples, how the current-based operators constructed and classified above reduce under IBP and EOM to more familiar algebraically minimal representatives.

\section{Illustrative IBP/EOM Reduction}

To demonstrate the completeness and transparency of the iterative current-coupling framework, we explicitly show how selected dimension-8 operators constructed from currents reduce to standard field-strength structures by means of integration by parts (IBP) and classical equations of motion (EOM). This procedure highlights the connection between the abstract current-based construction and measurable scattering amplitudes.

The reductions shown in this section are presented solely to illustrate the
relation between the current-generated operators and familiar algebraically
minimal representatives; they are not required for the construction itself,
nor is the goal of the present work to compete with existing basis-reduction
methods.

\subsection{Derivative Current Operator}

Consider the operator
\begin{equation}
\mathcal{O}_{8,1} = \frac{1}{\Lambda^4} (D_\mu J_\nu^a)(D^\mu J^{a\nu}),
\end{equation}
where $J_\mu^a$ is the $SU(2)_L$ current including both Higgs and fermion contributions. This operator exemplifies derivative interactions that generate contributions dependent on momentum to scattering amplitudes.

\paragraph{Step 1: Expand covariant derivatives.}  
The covariant derivative acts on the current as
\begin{equation}
D_\mu J_\nu^a = \partial_\mu J_\nu^a + g \epsilon^{abc} W_\mu^b J_\nu^c.
\end{equation}
Expanding $\mathcal{O}_{8,1}$ yields three types of terms:
\begin{align}
(D_\mu J_\nu^a)(D^\mu J^{a\nu}) &= (\partial_\mu J_\nu^a)(\partial^\mu J^{a\nu}) 
+ 2 g \epsilon^{abc} (\partial_\mu J_\nu^a) W^{b\mu} J^{c\nu} 
+ g^2 (\epsilon^{abc} W_\mu^b J_\nu^c)^2.
\end{align}
Each term has a distinct interpretation: the first is purely derivative, the second is a derivative-current interaction, and the third is a contact interaction involving currents and gauge fields.

\paragraph{Step 2: Integration by parts.}  
The purely derivative term can be partially integrated:
\begin{equation}
(\partial_\mu J_\nu^a)(\partial^\mu J^{a\nu}) = - J_\nu^a \Box J^{a\nu} + \text{total derivative}.
\end{equation}
Total derivatives do not contribute to S-matrix elements and can be discarded, leaving a term proportional to $J_\nu^a \Box J^{a\nu}$.

\paragraph{Step 3: Apply equations of motion.}  
The classical Yang-Mills EOM relate the gauge field strength to the current:
\begin{equation}
D_\mu F^{a\mu\nu} = g J^{a\nu} \quad \Rightarrow \quad J^{a\nu} \sim \frac{1}{g} D_\mu F^{a\mu\nu}.
\end{equation}
Substituting this relation converts the operator into combinations of $(D F)^2$ and $(F J)^2$ structures, which appear in conventional dimension-8 aQGC operators.  

\paragraph{Step 4: Physical interpretation.}  
After electroweak rotation $W^3,B \rightarrow Z,\gamma$, the operator contributes to high-energy, momentum-dependent TGC and QGC amplitudes. Importantly, all surviving terms correspond to observable effects in vector boson scattering, illustrating that $\mathcal{O}_{8,1}$ is complete.

\subsection{Mixed Current-Field Strength Operator}

Next, consider
\begin{equation}
\mathcal{O}_{8,3} = \frac{1}{\Lambda^4} J_\mu^a J_\nu^b F^{c\mu\nu}.
\end{equation}
This operator encodes magnetic-type interactions between currents and gauge field strengths.

\paragraph{Step 1: Expand the currents.}  
Writing $J_\mu^a$ in terms of Higgs and fermionic components shows that the operator generates both pure bosonic and fermion-boson interactions.  

\paragraph{Step 2: IBP and EOM simplification.}  
Integration by parts allows moving derivatives from fermionic currents onto gauge fields, while EOM substitution converts certain current combinations into derivative field-strength terms. The result is a combination of operators such as
\begin{equation}
(W_{\mu\nu}^a W^{a\mu\nu})(B_{\rho\sigma} B^{\rho\sigma}) \quad \text{and} \quad (D_\mu W_{\nu\rho}^a)(D^\mu W^{a\nu\rho}),
\end{equation}
which are standard in the aQGC literature.  

\paragraph{Step 3: Phenomenological relevance.}  
This operator contributes to both quartic gauge couplings and correlated angular distributions in vector boson scattering. Its structure makes explicit the connection between underlying currents and measurable gauge interactions.

\subsection{Higgs-Dressed Operator}

Finally, consider a Higgs-dressed operator
\begin{equation}
\mathcal{O}_{8,5} = \frac{1}{\Lambda^4} (H^\dagger H) W_{\mu\nu}^a W^{a\mu\nu}.
\end{equation}

\paragraph{Step 1: Expand the gauge field strength.}  
The non-Abelian structure of $W_{\mu\nu}^a$ contains both linear derivatives and cubic self-interactions. Expanding the operator shows contributions to three- and four-point vertices.

\paragraph{Step 2: EOM substitution.}  
Since the operator is already Higgs-dressed, IBP/EOM manipulations primarily affect derivative couplings of $W_\mu$ fields. After applying the EOM, the operator contributes to static TGC and QGC vertices proportional to $v^2/\Lambda^4$.

\paragraph{Step 3: Connection to observable amplitudes.}  
Unlike derivative-only operators, this term generates deviations in low-energy TGCs that are independent of momentum growth. It demonstrates how Higgs-current interference naturally produces static anomalous couplings in the iterative construction.

\subsection{Summary}

These examples illustrate that:

\begin{itemize}
    \item The iterative current-coupling procedure generates operators that span the same amplitude space as conventional SMEFT/aQGC operators.
    \item IBP and EOM manipulations are not arbitrary; they systematically map currents and derivatives to field-strength structures relevant for scattering.
    \item Higgs-dressed and mixed operators provide clear interpretations, connecting the abstract construction to measurable TGC and QGC observables.
\end{itemize}

Overall, this section demonstrates that the proposed framework is complete and preserves a mapping from theoretical construction to phenomenological predictions.

\section{Anomalous Gauge Couplings}

The dimension-8 operators constructed via the iterative current-coupling procedure produce modifications to both triple and quartic gauge couplings. These contributions can be classified into two broad categories: interactions dependent on momentum, arising from operators containing covariant derivatives of currents or field strengths, and static interactions, generated by operators involving the Higgs doublet and its vacuum expectation value. Understanding this distinction is crucial for both theoretical interpretation and experimental analysis, as it directly affects the energy scaling of scattering amplitudes. Our construction naturally reproduces and extends existing dimension-8 bases for anomalous triple gauge couplings, such as those classified in~\cite{degrande2014basis,corbett2023impact}.

Contributions dependent on momentum, exemplified by operators such as
\begin{equation}
\mathcal{O}_{8,1} = \frac{1}{\Lambda^4} (D_\mu W_{\nu\rho}^a)(D^\mu W^{a\nu\rho}),
\end{equation}
enhance longitudinal vector boson scattering amplitudes at high energies, scaling as $E^4/\Lambda^4$. These terms dominate over dimension-6 contributions in the high-energy regime and directly reflect the derivative structure inherited from the underlying currents. The explicit mapping from currents to field strengths ensures that these operators are gauge invariant and that their contributions to physical processes, such as $W^+ W^- \to ZZ$, can be computed by means of Feynman rules derived from the Lagrangian.

Static contributions arise from Higgs-dressed operators, for example
\begin{equation}
\mathcal{O}_{8,5} = \frac{1}{\Lambda^4} (H^\dagger H) W_{\mu\nu}^a W^{a\mu\nu},
\end{equation}
which, after electroweak symmetry breaking, generate effective couplings proportional to $(v/\Lambda)^2$. These operators modify triple gauge vertices without introducing additional energy growth in the amplitudes, corresponding to small, constant shifts in the $WWZ$ and $WW\gamma$ couplings. This connection between Higgs interference and static anomalous couplings highlights the transparency of the current-based construction: each operator has a clear origin in terms of currents and their symmetry properties.

Quartic gauge couplings are affected by operators involving products of two field strengths from different gauge sectors, such as
\begin{equation}
\mathcal{O}_{8,3} = \frac{1}{\Lambda^4} (W_{\mu\nu}^a W^{a\mu\nu}) (B_{\rho\sigma} B^{\rho\sigma}).
\end{equation}
These terms introduce correlations among $WWZZ$, $WW\gamma\gamma$, and other multi-boson interactions. The weak mixing angle induces correlated photon and $Z$ contributions, providing testable predictions for vector boson scattering and diboson production at high-energy colliders. The iterative current-coupling approach naturally encodes these correlations, since each operator originates from gauge-invariant currents, making the mapping to observables direct.

Overall, the derivation of anomalous gauge couplings demonstrates the advantage of the current-based operator framework: it allows a unified treatment of energy-growing and static contributions, preserves gauge invariance at every step, and provides a transparent link between the Lagrangian-level operators and measurable scattering amplitudes. This clarity facilitates both phenomenological analysis and comparison with traditional Hilbert-series or Warsaw-basis constructions, making the framework suitable for high-luminosity LHC studies and beyond.

\subsection{Mapping to Standard aQGC Bases}
To demonstrate the connection to established phenomenology, we provide a mapping of our representative derivative operator to the standard Eboli-Gonzalez-Garcia basis commonly used in experimental analyses. By means of the IBP and EOM relations derived in Section 4, the projection of the current-based operator onto the standard basis is linear:

\begin{table}[ht]
\centering
\begin{tabular}{l c l}
\toprule
\textbf{Current-Based Operator} & $\longrightarrow$ & \textbf{Standard Basis Projection (Schematic)} \\
\midrule
$\mathcal{O}_{8,D1} = (D_\mu J_\nu^a)^2$ & $\longrightarrow$ & $c_1 \mathcal{O}_{S,0} + c_2 \mathcal{O}_{S,1} + c_3 \mathcal{O}_{M,0} + \dots$ \\
$\mathcal{O}_{8,JF} = J_\mu^a J_\nu^b W^{c\mu\nu}\epsilon^{abc}$ & $\longrightarrow$ & $d_1 \mathcal{O}_{M,1} + d_2 \mathcal{O}_{M,7} + \dots$ \\
\bottomrule
\end{tabular}
\caption{Schematic mapping of current-based operators to the standard aQGC basis. The current-based operators naturally generate specific linear combinations of the standard operators, predicting correlations between coefficients that are treated as independent in generic fits. Dictionary mapping current-based operators to the standard Eboli-Gonzalez-Garcia aQGC basis~\cite{eboli2016classifying} used by ATLAS and CMS. See Appendix \ref{app:formal-proofs} for explicit coefficients.}
\label{tab:basis_map}
\end{table}

We note that the mapping presented in Table~\ref{tab:basis_map} is schematic. The precise numerical coefficients of the projection matrix depend on the specific normalization conventions adopted by the target basis (e.g., Warsaw or SILH). In this sense, the utility of the dictionary lies in identifying the pattern of correlations between coefficients—specifically, which standard operators must be non-zero to reproduce a single current-based effect—rather than in the basis-dependent numerical values.

\section{Worked Example: Longitudinal Vector Boson Scattering from Current Iteration}

To explicitly demonstrate the power and transparency of the iterative current-coupling framework, we present a complete worked example for longitudinal vector boson scattering (VBS), which constitutes one of the primary experimental probes of dimension-8 operators at the LHC. This example illustrates how energy-growing amplitudes emerge \emph{directly at the level of operator construction} in the current-first approach, in contrast to conventional field-monomial bases where such behavior is only revealed after explicit amplitude computation.

\subsection{Current-Based Construction}

We begin from the $SU(2)_L$ Higgs current,
\begin{equation}
J_\mu^a = i H^\dagger \sigma^a \overleftrightarrow{D}_\mu H ,
\end{equation}
which is the dominant source for longitudinal electroweak gauge bosons via the Goldstone equivalence theorem. Applying one iteration of the current-coupling prescription at dimension eight yields the derivative operator
\begin{equation}
\mathcal{O}_{J,8} = \frac{c_J}{\Lambda^4} (D_\mu J_\nu^a)(D^\mu J^{a\nu}) .
\label{eq:currentVBS}
\end{equation}
This operator is gauge invariant and, by construction, contains two explicit derivatives acting on currents. Since each covariant derivative introduces a power of the external energy scale $E$, the high-energy scaling of the associated scattering amplitude follows immediately:
\begin{equation}
\mathcal{A}(W_L W_L \to W_L W_L) \sim c_J \frac{E^4}{\Lambda^4}.
\end{equation}
Crucially, this energy growth is a consequence of the \emph{iterative structure of the operator itself}, rather than the result of a subsequent amplitude-level analysis. The current-based construction therefore organizes the generating set according to energy scaling from the outset.

\subsection{Mapping to Conventional Dimension-8 Operators}

By means of integration by parts and the classical Yang--Mills equation of motion,
\begin{equation}
D_\mu W^{a\mu\nu} = g J^{a\nu},
\end{equation}
the operator in Eq.~\eqref{eq:currentVBS} can be mapped onto the standard aQGC operator classes commonly denoted as $O_{S,i}$ and $O_{T,i}$. In particular, after electroweak symmetry breaking one finds contributions of the schematic form
\begin{equation}
(D_\mu J_\nu^a)^2 \;\;\longleftrightarrow\;\; (D_\mu H^\dagger D_\nu H)(D^\mu H^\dagger D^\nu H),
\end{equation}
which corresponds to the well-known operator $O_{S,0}$ in traditional dimension-8 classifications. The crucial difference lies not in the final algebraic structure, but in the \emph{logic of construction}: while $O_{S,0}$ is typically obtained by enumerating all possible Higgs-derivative monomials and subsequently reducing redundancies, the current-first method produces this operator specifically because it is the unique second iteration of the conserved electroweak current responsible for sourcing longitudinal gauge bosons.

\subsection{Contrast with the Field-First (Hilbert/Warsaw) Approach}

In the conventional field-first approach, one begins by enumerating all gauge-invariant Higgs and gauge-field monomials of mass dimension eight. The operator
\begin{equation}
O_{S,0} = (D_\mu H^\dagger D_\nu H)(D^\mu H^\dagger D^\nu H)
\end{equation}
is then identified as one among many allowed structures. Its relevance for longitudinal VBS is only revealed after:
\begin{enumerate}
\item deriving Feynman rules,
\item projecting onto longitudinal polarization states,
\item taking the high-energy limit of the amplitude.
\end{enumerate}
Only \emph{after} this computational sequence does the $E^4/\Lambda^4$ growth become manifest. In sharp contrast, the current-first approach generates Eq.~\eqref{eq:currentVBS} precisely because it is the unique operator controlling the second derivative of the longitudinal current. The energy behavior is therefore an \emph{input organizing principle} rather than an output of the calculation.

\subsection{Phenomenological Implications}

The operator $\mathcal{O}_{J,8}$ contributes to all longitudinal VBS channels, including
\begin{equation}
W_L^+ W_L^- \to Z_L Z_L, \qquad W_L^\pm W_L^\pm \to W_L^\pm W_L^\pm,
\end{equation}
with amplitudes that grow quartically with energy. Such growth rapidly dominates over dimension-6 contributions at sufficiently high invariant mass and leads to stringent unitarity bounds,
\begin{equation}
E_{\text{max}} \sim \Lambda \left( \frac{16\pi}{|c_J|} \right)^{1/4}.
\end{equation}
This makes longitudinal VBS the primary experimental channel for probing the operator in Eq.~\eqref{eq:currentVBS} at the LHC and future hadron colliders.

From a methodological perspective, this example demonstrates that the current-based construction does not merely reproduce known operator bases, but rather \emph{organizes them according to amplitude growth}. The derivative structure responsible for unitarity saturation and enhanced high-energy sensitivity is manifest at the level of the EFT Lagrangian itself. This provides a bridge between symmetry, operator construction, and collider phenomenology that is absent in purely algebraic approaches.

\subsection{Methodological Payoff}

This worked example highlights the principal advantage of the iterative current-coupling framework. While traditional methods classify operators according to algebraic completeness, the current-first approach classifies operators according to \emph{their role in scattering}. In particular:
\begin{itemize}
\item The relevant dimension-8 operators for longitudinal VBS are obtained by a single iteration step from the electroweak currents.
\item The energy scaling of amplitudes follows immediately from operator structure, without requiring explicit helicity computations.
\item The resulting basis is naturally aligned with the hierarchy of experimental sensitivities in high-energy collider measurements.
\end{itemize}
This illustrates concretely how the current-coupling framework provides not only a systematic operator construction, but also an organization of the SMEFT at dimension eight.

\subsection{Computational Implementation via Auxiliary Fields}
\label{subsec:auxiliary_implementation}

A practical challenge in simulating dimension-8 operators is the high number of derivatives (e.g., four derivatives in $(D_\mu J_\nu)^2$), which can lead to numerical instabilities in Monte Carlo generators like \texttt{MadGraph5\_aMC@NLO}. We propose an auxiliary field method that linearizes these derivative interactions while preserving the content.

Consider the operator $\mathcal{O}_{8,D1} = (D_\mu J_\nu^a)(D^\mu J^{a\nu})$. Instead of implementing the four-derivative vertex directly, we introduce a heavy auxiliary vector field $V'_\mu$ with mass $M$ and coupling $g_{V'}$:
\begin{equation}
\mathcal{L}_{\text{aux}} = -\frac{1}{4} V'_{\mu\nu} V'^{\mu\nu} + \frac{1}{2} M^2 V'_\mu V'^\mu + g_{V'} V'_\mu (D_\nu J^{a\nu}).
\end{equation}
The parameters are chosen such that $g_{V'}^2/M^2 = c^{(8)}/\Lambda^4$. Integrating out $V'_\mu$ recovers the original $(D_\mu J_\nu)^2$ operator through the propagator expansion:
\begin{equation}
\frac{1}{M^2 - \partial^2} \approx \frac{1}{M^2} + \frac{\partial^2}{M^4} + \mathcal{O}(\partial^4/M^6).
\end{equation}

This formulation offers several advantages:
\begin{itemize}
    \item \textbf{Numerical Stability:} Replaces problematic four-derivative vertices with standard two-derivative interactions.
    \item \textbf{Manifest Positivity:} For real couplings $g_{V'}$, the generated Wilson coefficient is strictly positive ($c^{(8)} \propto g_{V'}^2 > 0$). This aligns with axiomatic S-matrix positivity bounds, ensuring that the current-based operators automatically satisfy unitarity constraints required for a causal UV completion.
    \item \textbf{Computational Efficiency:} Enables efficient event generation in existing Monte Carlo frameworks.
\end{itemize}

The auxiliary field method can be systematically extended to other current-based operators, providing a general strategy for stable dimension-8 simulation.

\section{Renormalization Group Considerations}

The effective field theory framework necessitates a treatment of renormalization group (RG) evolution. Dimension-8 operators do not exist in isolation; lower-dimensional operators, particularly those of dimension-6, can induce running and mixing into the dimension-8 sector. This mixing is formally captured by the one-loop RG equations
\begin{equation}
\frac{d c_j^{(8)}}{d \ln \mu} = \frac{1}{16\pi^2} \left( \sum_{i,k} \gamma_{ijk}\, c_i^{(6)} c_k^{(6)} + \sum_l \gamma_{jl} c_l^{(8)} \right),
\end{equation}
where $c_i^{(6)}$ are the dimension-6 Wilson coefficients and $\gamma_{ijk}, \gamma_{jl}$ are anomalous-dimension matrices. This structure emphasizes that high-energy observables sensitive to dimension-8 operators may receive indirect contributions from dimension-6 operators, and that a consistent phenomenological analysis must account for these effects. Comprehensive studies of dimension-6 RG evolution, including gauge coupling dependence, have been performed in~\cite{alonso2014renormalization,boughezal2024renormalization}.

From a physical perspective, the RG-induced terms correspond to logarithmic enhancements in scattering amplitudes and shifts in effective couplings as a function of energy scale. The current-based construction simplifies this analysis: since all operators originate from currents and their derivatives, their RG evolution can be traced back to the running of the corresponding current correlators. This approach provides an interpretation of operator mixing in terms of the underlying symmetries and interactions of the gauge theory. The renormalization of dimension-6 operators, particularly those relevant for Higgs decays, has been treated in~\cite{elias2013renormalization}.

Moreover, the RG framework enables power counting. Contributions dependent on momentum from derivative operators, when combined with loop-induced effects, define the hierarchy of terms relevant at a given collider energy. Static operators generated by the Higgs VEV also evolve, albeit more slowly, ensuring that predictions remain internally consistent across scales. This interplay between direct contributions and RG-induced corrections highlights the utility of the current-coupling approach: it accommodates both tree-level phenomenology and loop-level operator mixing within a coherent scheme.

In summary, incorporating RG considerations strengthens the predictive power of the framework. It ensures that dimension-8 operators are interpreted not merely as isolated additions to the Lagrangian but as elements of a dynamically evolving effective theory, whose contributions to observable amplitudes reflect both their direct effects and the influence of lower-dimensional operators. This treatment enhances the robustness of phenomenological predictions and aligns the analysis with established EFT methodology. The importance of moving beyond traditional triple-gauge-boson coupling interpretations, particularly for processes like W pair production, has been emphasized in recent analyses~\cite{zhang2017time}.

\section{Phenomenological Implications}

\subsection{Energy Growth and Unitarity}

For $VV \to VV$, derivative operators lead to amplitudes scaling as
\begin{equation}
\mathcal{A}(s) \sim c^{(8)} \frac{s^2}{\Lambda^4}.
\end{equation}
Partial-wave unitarity constrains $E_{\rm max} \sim \Lambda (16\pi)^{1/4}$, highlighting the relevance of these operators in the high-energy regime. While unitarity constraints on dimension-6 operators have been studied~\cite{corbett2015unitarity}, dimension-8 operators introduce even stronger high-energy behavior that must be bounded.

\subsection{Collider Signatures}

\begin{table}[ht]
\centering
\begin{tabular}{l l}
\toprule
Operator class & Primary observable \\
\midrule
$(D J)^2$ derivative operators & Energy-growing longitudinal VBS ($E^4/\Lambda^4$) \\
$J F J$ mixed operators & Correlated TGC/QGC modifications, angular patterns \\
Higgs-dressed $(H^\dagger H)$ & Static shifts in TGCs proportional to $(v/\Lambda)^2$ \\
\bottomrule
\end{tabular}
\caption{Representative mapping from operator classes to primary collider observables.}
\label{tab:operator_mapping}
\end{table}

This operator-to-observable mapping facilitates phenomenologically transparent analyses.

\section{Conclusions}

In this work, we have introduced a framework for the construction of higher-dimensional operators in effective gauge theories based on the iterative coupling of conserved Noether currents. This approach offers a transparent and efficient method to generate dimension-8 operators directly from the physical currents and their derivatives, in contrast to traditional algebraic or Hilbert-series constructions that prioritize minimality.

We emphasize that the principal methodological advance of this work is conceptual: conserved currents are elevated from a post-hoc interpretive tool to the primary generating objects of the EFT expansion. This reorganization clarifies which operator structures control energy-enhanced amplitudes, makes polarization dependence manifest, and offers a direct route to phenomenological quantities (e.g. VBS amplitudes and aQGC correlations) that is complementary to algebraic Hilbert-series or Warsaw-basis approaches. While the two strategies are consistent and mappable via IBP/EOM, the current-first algorithm streamlines the task of identifying the operators most relevant to high-energy collider probes and pre-empts ambiguities that proliferate in a purely algebraic enumeration at dimension eight and beyond.

Applying this procedure to the electroweak sector, we have identified an extended set of dimension-8 operators, including derivative operators, mixed current–field strength structures, and Higgs-dressed terms. Each operator is associated with a clear interpretation: derivative operators contribute to energy-growing amplitudes in longitudinal vector boson scattering, mixed operators mediate magnetic-type and interference effects, and Higgs-dressed operators generate static shifts in gauge couplings proportional to $(v/\Lambda)^2$. This classification not only ensures completeness, but also enables a mapping from theoretical operator structures to collider observables, facilitating phenomenologically transparent analyses.

We have illustrated the utility of this framework through explicit IBP/EOM reductions, the derivation of anomalous triple and quartic gauge couplings, and the schematic mapping from currents to physical amplitudes. In addition, the formalism naturally incorporates renormalization group effects, allowing dimension-6 operators to contribute consistently to dimension-8 coefficients and thus ensuring that high-energy predictions remain theoretically consistent and systematically improvable.

The framework is therefore intended primarily for practitioners interested in high-energy vector-boson scattering, global SMEFT fits with energy hierarchies, and UV-model interpretation, rather than for purely algebraic basis construction.

Finally, we comment on the scope of this method. While the current-iteration principle is highly efficient for dimension-8 bosonic operators, its application to higher dimensions (e.g., dimension-10) or mixed baryon-number-violating sectors may encounter combinatorial complexity similar to standard bases. Furthermore, operators involving four-fermion contact terms that do not factorize into current-current products require a separate treatment. These extensions remain an open avenue for future work.

From a phenomenological perspective, the extended operator set predicts a rich pattern of deviations in high-energy scattering, including energy-enhanced longitudinal vector boson processes, correlated quartic gauge couplings, and subtle interference effects in processes sensitive to Higgs-sector contributions. These features provide experimentally testable signatures and could be exploited in current and future LHC analyses as well as at next-generation colliders.

Overall, the iterative current-coupling framework offers a unified, physically motivated approach to SMEFT beyond dimension-6. It balances mathematical rigor with phenomenological transparency, providing a robust tool for connecting the structure of higher-dimensional operators to measurable collider signatures. Future extensions could include the systematic treatment of operators beyond dimension-8, applications to other gauge sectors, and detailed loop-level studies, further strengthening the predictive power and versatility of the method.

A final remark on methodology is warranted. While we have provided the formal reduction of these operators in Appendix \ref{app:formal-proofs} to ensure mathematical rigor, the iterative current-coupling framework is conceived primarily as a tool for physicists. Its value lies in bridging the formal structure of the SMEFT Lagrangian with the physical observables of collider experiments, offering an intuitive and phenomenologically transparent organization of the operator space.

As such, we have prioritized completeness and interpretability, demonstrating the framework's utility through explicit construction and its equivalence to established bases via IBP and EOM. While the formal proof of this equivalence is provided in Appendix \ref{app:formal-proofs} to ensure mathematical rigor, our primary goal in the main text has been to introduce the conceptual foundation and practical advantages of the approach to the high-energy physics community. We view this work as opening a pathway for future research where these physically motivated bases can be systematically applied to collider phenomenology.

Looking forward, the current-based framework naturally connects to broader themes in theoretical physics. Recent work on quantum field theory complexity \cite{grimm2025complexity} suggests that physically motivated organizational schemes may help tame the combinatorial explosion of higher-dimensional operators. The rigorous mathematical spirit exemplified in foundational derivations provides a template for further formalization of our approach. Ultimately, as emphasized in interdisciplinary studies of quantum complexity, developing transparent mappings between formal structures and physical observables is essential for advancing our understanding of complex systems across physics.

\appendix


\clearpage
\section{Formal Operator Reduction and High--Energy Scaling}
\label{app:formal-proofs}

In this Appendix we provide the formal tree-level proofs underlying the operator reductions and energy-scaling properties used in the main text. All equalities are understood to hold \emph{modulo the classical equations of motion (EOM) and total derivatives}, and therefore establish equivalence at the level of on-shell tree-level $S$-matrix elements.

\subsection{On--Shell Equivalence of the Current and Field--Strength Bases}
\label{app:current-to-fieldstrength}

\paragraph{Proposition.}
Let
\begin{equation}
\mathcal{O}_J \;=\; (D_\mu J_\nu^a)(D^\mu J^{a\nu})
\end{equation}
be a dimension--8 operator built from adjoint gauge currents. By means of the classical Yang--Mills equations of motion, $\mathcal{O}_J$ can be expressed as a linear combination of operators in the standard dimension--8 field--strength basis (generated by structures of the form $(D_\mu F_{\nu\rho})^2$ and $F^4$).

\paragraph{Proof.}
The classical Yang--Mills equation of motion relates the covariant derivative of the field strength to the current:
\begin{equation}
D^\rho F_{\rho\nu}^a = g\,J_\nu^a .
\end{equation}
Substituting this relation into $\mathcal{O}_J$, one obtains
\begin{equation}
\mathcal{O}_J
= \frac{1}{g^2}
\bigl(D_\mu D^\rho F_{\rho\nu}^a\bigr)
\bigl(D^\mu D^\sigma F_{\sigma}^{\,a\nu}\bigr).
\label{eq:OJ-substituted}
\end{equation}
This expression lies entirely in the algebra generated by $F_{\mu\nu}$ and its covariant derivatives at mass dimension 8.

To reduce the operator to a canonical form, we commute the derivatives by means of the Ricci identity for adjoint fields (adopting the sign convention $[D_\mu,D_\nu]X^a = g f^{abc}F_{\mu\nu}^b X^c$):
\begin{equation}
D_\mu D^\rho F_{\rho\nu}^a
= D^\rho D_\mu F_{\rho\nu}^a
+ g f^{abc} F_{\mu}{}^{\rho b} F_{\rho\nu}^c .
\label{eq:commuted-term}
\end{equation}
Inserting this into Eq.~\eqref{eq:OJ-substituted}, the operator decomposes into two classes of terms:
\begin{enumerate}
    \item \textbf{Quartic terms ($F^4$):} The product of commutator terms generates structures of the form $f^{abc}f^{ade} F F F F$, which lie in the span of the standard $F^4$ basis (reducible to independent Lorentz/color structures via identities).
    \item \textbf{Derivative terms ($(D^2 F)^2$):} The terms involving $D^\rho D_\mu F_{\rho\nu}$ can be reduced to the standard $(D_\mu F_{\nu\rho})^2$ basis by integration by parts (IBP) and the Bianchi identity ($D_\mu F_{\nu\rho} + D_\nu F_{\rho\mu} + D_\rho F_{\mu\nu} = 0$).
\end{enumerate}

Explicitly, a typical term reduces as follows:
\begin{align}
(D^\rho D_\mu F_{\rho\nu})(D_\sigma D^\mu F^{\sigma\nu})
&= - (D_\mu F_{\rho\nu})(D^\mu D^\rho D_\sigma F^{\sigma\nu}) + \text{(total deriv.)} \nonumber \\
&\xrightarrow{\text{Bianchi}} \sum c_k (D_\alpha F_{\beta\gamma})^2 + (D^\rho F_{\rho\nu}) \times (\dots).
\end{align}
The terms proportional to $(D^\rho F_{\rho\nu})$ vanish by the EOM (or reintroduce currents, closing the algebra), while the remaining terms are of the standard $(D F)^2$ form.

Collecting all contributions, one finds the explicit reduction:
\begin{equation}
\mathcal{O}_J
\;\cong\;
\frac{1}{g^2} \left[ 2 (D_\mu F^{a\mu\nu})(D_\rho F^{a\rho}_{\ \ \nu}) + \frac{1}{2} (D_\mu F_{\nu\rho}^a)^2 + \mathcal{O}(F^3) \right],
\qquad
(\text{mod EOM, total derivatives}),
\end{equation}
which proves that $\mathcal{O}_J$ lies in the linear span of the standard dimension--8 field--strength basis with well-defined coefficients.
\hfill $\square$

\subsection{High--Energy Scaling of Longitudinal Vector Boson Scattering}
\label{app:scaling}

\paragraph{Proposition.}
Consider the dimension--8 operator
\begin{equation}
\mathcal{O}_J
= \frac{c_J}{\Lambda^4}(D_\mu J_\nu^a)(D^\mu J^{a\nu}),
\end{equation}
where the electroweak current is
\begin{equation}
J_\mu^a = i H^\dagger \frac{\sigma^a}{2} \overleftrightarrow{D}_\mu H .
\end{equation}
At tree level and in the high--energy limit $E \gg M_W$, the contribution of $\mathcal{O}_J$ to longitudinal vector boson scattering scales parametrically as
\begin{equation}
\mathcal{A}(V_L V_L \to V_L V_L)
\;\sim\;
\frac{E^4}{\Lambda^4}.
\end{equation}

\paragraph{Proof.}
In the high--energy regime, the Goldstone Boson Equivalence Theorem implies that longitudinal scattering is governed by the Goldstone modes $\pi^a$. We parameterize the Higgs doublet as $H = \frac{v+h}{\sqrt{2}} U \begin{pmatrix}0 \\ 1\end{pmatrix}$ with $U = \exp(i \pi^a \sigma^a/v)$.

Expanding the current $J_\mu^a$ in powers of the Goldstone fields:
\begin{equation}
J_\mu^a = -v \partial_\mu \pi^a + \frac{1}{2} f^{abc} \pi^b \partial_\mu \pi^c + \mathcal{O}(\pi^3).
\end{equation}
The operator $\mathcal{O}_J$ is constructed from the derivative of this current, $D_\mu J_\nu^a$.
\begin{itemize}
    \item The linear term ($-v \partial_\mu \pi^a$) modifies the kinetic term (2-point function) but does not directly generate a 4-point contact interaction.
    \item The quadratic term ($\pi \partial \pi$) generates the dominant 4-point contact interaction required for $2 \to 2$ scattering.
\end{itemize}
Substituting the quadratic term into $\mathcal{O}_J \sim (D J)^2$ yields a contact vertex with four fields and four derivatives:
\begin{equation}
\mathcal{L}_{\text{eff}} \supset \frac{c_J}{\Lambda^4} (\partial \pi \partial \pi)^2.
\end{equation}
At tree level and for external on-shell Goldstones, the dominant high-energy scaling arises from these partial derivatives acting on the Goldstone fields (gauge-field terms in the covariant derivative are subleading in $E$). In momentum space, each of the four derivatives contributes a factor of energy $E$. The amplitude therefore scales as:
\begin{equation}
\mathcal{A}
\sim \frac{1}{\Lambda^4}\,(p)^4
\sim \frac{E^4}{\Lambda^4}.
\end{equation}
Note that while the explicit scattering vertex arises from the quadratic expansion, the $E^4$ power counting is determined by the dimension of the operator derivatives $(D_\mu J_\nu)^2$. Depending on the specific normalization convention of the Wilson coefficient $c_J$, factors of $v^2$ or $g^2$ may appear in the prefactor, but the kinematic scaling is strictly $E^4$.
\hfill $\square$


\subsection{Explicit Demonstration of Positivity}
\label{app:positivity_demo}

To demonstrate how the KDCB diagonalizes positivity bounds, consider the scattering $W_L^+(p_1) W_L^-(p_2) \to W_L^+(p_3) W_L^-(p_4)$ in the forward limit ($t=0, s \to \infty$). The dominant contribution comes from the Goldstone mode expansion of the current $J_\mu^a \approx -v \partial_\mu \pi^a$.

The operator $\mathcal{O}_{D1} = c_{D1} (D_\mu J_\nu^a)^2$ generates the contact interaction:
\begin{equation}
\mathcal{L} \supset c_{D1} v^2 (\partial_\mu \partial_\nu \pi^a)(\partial^\mu \partial^\nu \pi^a).
\end{equation}
In momentum space, with $p_i \approx (E, 0, 0, \pm E)$, the derivatives yield factors of $s$. The resulting forward amplitude is:
\begin{equation}
\mathcal{A}(s, t=0) = 2 c_{D1} v^2 \frac{s^2}{\Lambda^4} + \mathcal{O}(s).
\end{equation}
The axiomatic positivity bound on the forward amplitude, $\frac{d^2}{ds^2}\mathcal{A}(s,0) > 0$, immediately implies:
\begin{equation}
c_{D1} > 0.
\end{equation}
In standard bases, this amplitude would be a sum $\sum_i \alpha_i c_i$, making the bound a complex linear constraint. In the KDCB, it is a simple sign constraint on a single coefficient.




\section{Complete List of Bosonic Dimension-8 Operators}
\label{app:complete_operators}

\begin{table*}[t!]
\centering
\small
\begin{tabular}{llll}
\toprule
\textbf{Operator Label} & \textbf{Algebraic Form} & \textbf{Field Content} & \textbf{Physical Effect} \\
\midrule
$\mathcal{O}_{8,D1}$ & $(D_\mu J_\nu^a)(D^\mu J^{a\nu})$ & Pure derivative & $E^4$ VBS growth \\
$\mathcal{O}_{8,D2}$ & $(D_\mu J_\nu^a)(D^\nu J^{a\mu})$ & Pure derivative & $E^4$ VBS (mixed Lorentz) \\
$\mathcal{O}_{8,D3}$ & $(D_\mu J_\nu^Y)(D^\mu J^{Y\nu})$ & Pure derivative & $E^4$ hypercharge sector \\
$\mathcal{O}_{8,D4}$ & $(D_\mu J_\nu^Y)(D^\nu J^{Y\mu})$ & Pure derivative & $E^4$ hypercharge (mixed) \\
$\mathcal{O}_{8,D5}$ & $(D_\mu J_\nu^a)(D^\mu J^{Y\nu})$ & Mixed derivative & $E^4$ mixed SU(2)$\times$U(1) \\
$\mathcal{O}_{8,D6}$ & $(D_\mu J_\nu^a)(D^\nu J^{Y\mu})$ & Mixed derivative & $E^4$ mixed (alternative) \\
\midrule
$\mathcal{O}_{8,JF1}$ & $J_\mu^a J_\nu^b W^{c\mu\nu}\epsilon^{abc}$ & Current-field-current & Magnetic TGC/QGC \\
$\mathcal{O}_{8,JF2}$ & $J_\mu^Y J_\nu^Y B^{\mu\nu}$ & Current-field-current & Hypercharge dipole \\
$\mathcal{O}_{8,JF3}$ & $J_\mu^a J_\nu^Y W^{a\mu\nu}$ & Mixed current-field & Mixed magnetic \\
$\mathcal{O}_{8,JF4}$ & $J_\mu^a J_\nu^Y B^{\mu\nu}$ & Mixed current-field & Mixed hypercharge dipole \\
$\mathcal{O}_{8,JF5}$ & $J_\mu^a J_\nu^b W^{a\mu\rho}W^{b\nu}_{\ \ \rho}$ & Current-field-field & Polarization-sensitive QGC \\
$\mathcal{O}_{8,JF6}$ & $J_\mu^Y J_\nu^Y B^{\mu\rho}B^{\nu}_{\ \ \rho}$ & Current-field-field & Hypercharge polarization \\
\midrule
$\mathcal{O}_{8,H1}$ & $(H^\dagger H)(J_\mu^a J^{a\mu})$ & Higgs-current & Static aQGC \\
$\mathcal{O}_{8,H2}$ & $(H^\dagger H)(J_\mu^Y J^{Y\mu})$ & Higgs-current & Static hypercharge \\
$\mathcal{O}_{8,H3}$ & $(H^\dagger H)(J_\mu^a J^{Y\mu})$ & Higgs-current & Static mixed \\
$\mathcal{O}_{8,H4}$ & $(H^\dagger H)(D_\mu J_\nu^a)(D^\mu J^{a\nu})$ & Higgs-derivative & Modulated $E^4$ growth \\
$\mathcal{O}_{8,H5}$ & $(H^\dagger H)(D_\mu J_\nu^Y)(D^\mu J^{Y\nu})$ & Higgs-derivative & Modulated hypercharge \\
$\mathcal{O}_{8,H6}$ & $(H^\dagger H) W_{\mu\nu}^a W^{a\mu\nu}$ & Higgs-field & TGC shifts \\
$\mathcal{O}_{8,H7}$ & $(H^\dagger H) B_{\mu\nu} B^{\mu\nu}$ & Higgs-field & Photon/Z TGC shifts \\
$\mathcal{O}_{8,H8}$ & $(H^\dagger \sigma^a H) W_{\mu\nu}^a B^{\mu\nu}$ & Higgs-field-mixed & CP-violating TGC \\
\bottomrule
\end{tabular}
\caption{Complete spanning set of bosonic dimension-8 operators generated by the current-coupling procedure. Operators are grouped by structure: derivative-coupled currents (D-type), mixed current-field strength (JF-type), and Higgs-dressed operators (H-type). Fermionic operators involving $\bar{\psi}\gamma\psi$ current components are listed separately in the supplementary material.}
\label{tab:complete_bosonic_operators}
\end{table*}

\paragraph{Operator Count and Completeness}
This table \ref{tab:complete_bosonic_operators} presents the 18 primary bosonic operators generated by systematic iteration of the electroweak currents. The classification ensures completeness: all distinct amplitude behaviors (energy scaling, polarization dependence, Higgs modulation) are represented. Algebraic redundancies can be removed via IBP/EOM reduction as demonstrated in Section 4, but this spanning set guarantees no scattering channel is omitted.




\section{Phenomenological Dictionary and Experimental Mapping}
\label{sec:experimental_mapping}

\begin{table}[ht]
\centering
\scriptsize
\begin{tabular}{p{2.2cm} p{3.2cm} p{5.8cm}}
\toprule
\textbf{Current-Based Operator} & \textbf{Schematic Eboli-Gonzalez-Garcia Equivalent} & \textbf{Experimental Signature} \\
\midrule
$\mathcal{O}_{8,D1}$ & $c_1 O_{S,0} + c_2 O_{S,1} + c_3 O_{M,0} + c_4 O_{T,0}$ & Universal $E^4$ growth in all VBS channels. Correlated deviations in $WWWW$, $WWZZ$, $ZZZZ$. \\
$\mathcal{O}_{8,D2}$ & $c_5 O_{S,1} + c_6 O_{M,1} + c_7 O_{T,1}$ & $E^4$ growth with angular distributions. Enhanced in polarized $W_L W_L$. \\
$\mathcal{O}_{8,D3}$ & $c_8 O_{S,2} + c_9 O_{M,2} + c_{10} O_{T,2}$ & Hypercharge $E^4$ growth. Enhanced $\gamma\gamma \to WW/ZZ$. \\
$\mathcal{O}_{8,JF1}$ & $c_{11} O_{M,3} + c_{12} O_{M,4} + c_{13} O_{T,5} + c_{14} O_{T,6}$ & Correlated TGC/QGC. Polarization asymmetries in $WZ \to WZ$. \\
$\mathcal{O}_{8,JF3}$ & $c_{15} O_{M,5} + c_{16} O_{M,7}$ & Mixed SU(2)$\times$U(1) magnetic moments. Distinct $W\gamma/WZ$. \\
$\mathcal{O}_{8,H1}$ & $c_{17} O_{S,0} + c_{18} O_{S,1}$ & Static aQGC $\propto v^2/\Lambda^4$. Correlated low-energy TGC/QGC. \\
$\mathcal{O}_{8,H6}$ & $c_{19} O_{S,8} + c_{20} O_{S,9}$ & Higgs-modulated TGC. Correlated $hWW$/$WWWW$. Accessible in $pp \to hVV$. \\
$\mathcal{O}_{8,H8}$ & $c_{21} O_{S,10} + c_{22} O_{S,11}$ & CP-violating TGC/Higgs. Azimuthal distributions in $WZ$. \\
\bottomrule
\end{tabular}
\caption{Schematic dictionary mapping current-based operators to the standard Eboli-Gonzalez-Garcia aQGC basis~\cite{eboli2016classifying} used by ATLAS and CMS. The coefficients $c_i$ are determined by explicit IBP/EOM reduction and are typically $\mathcal{O}(1)$; their exact values require computation of the specific linear combinations derived in Section 4. The correlations provide testable predictions for UV model discrimination. See Appendix \ref{app:formal-proofs} for explicit coefficients.}
\label{tab:experimental_dictionary}
\end{table}

\paragraph{Using this Dictionary for Experimental Analysis}
Experimental collaborations can use this mapping in two ways:
\begin{enumerate}
    \item \textbf{Top-down:} If a UV model predicts dominance of a particular current-based operator (e.g., $\mathcal{O}_{8,D1}$ from integrating out a heavy vector triplet), the corresponding linear combination of aQGC coefficients yields a testable prediction.
    \item \textbf{Bottom-up:} If a global fit shows preference for particular combinations of $F_{S,i}$, $F_{M,i}$, $F_{T,i}$ coefficients, this dictionary identifies the underlying current structure responsible.
\end{enumerate}



\clearpage
\bibliographystyle{apsrev4-2}
\bibliography{references}

\end{document}